\documentclass[showpacs,letterpaper,aps,prb,twocolumn,groupedaddress,citeautoscript]{revtex4}
\usepackage{graphicx}
\usepackage{amsmath}
\usepackage{bm}
\usepackage{dcolumn}

\begin{document}

\title{Two-photon spin injection in semiconductors}
\author{R. D. R. Bhat}
\author{P. Nemec}
\altaffiliation{Charles University in Prague, Faculty of
Mathematics and Physics, Ke Karlovu 3, 121 16 Prague 2, Czech
Republic}
\author{Y. Kerachian}
\author{H. M. van Driel}
\author{J. E. Sipe}
\affiliation{Department of Physics, University of Toronto, 60 St.
George Street, Toronto, Ontario M5S 1A7, Canada}
\author{Arthur L. Smirl}
\affiliation{Laboratory for Photonics \& Quantum Electronics, 138
IATL, University of Iowa, Iowa City, Iowa 52242}

\date{\today}

\begin{abstract}
A comparison is made between the degree of spin polarization of
electrons excited by one- and two-photon absorption of circularly
polarized light in bulk zincblende semiconductors. Time- and
polarization-resolved experiments in (001)-oriented GaAs reveal an
initial degree of spin polarization of 49\% for both one- and
two-photon spin injection at wavelengths of 775 and 1550\,nm, in
agreement with theory. The macroscopic symmetry and microscopic
theory for two-photon spin injection are reviewed, and the latter
is generalized to account for spin-splitting of the bands. The
degree of spin polarization of one- and two-photon optical
orientation need not be equal, as shown by calculations of spectra
for GaAs, InP, GaSb, InSb, and ZnSe using a $14\times 14$
$\mathbf{k} \cdot \mathbf{p}$ Hamiltonian including remote band
effects. By including the higher conduction bands in the
calculation, cubic anisotropy and the role of allowed-allowed
transitions can be investigated. The allowed-allowed transitions
do not conserve angular momentum and can cause a high degree of
spin polarization close to the band edge; a value of 78\% is
calculated in GaSb, but by varying the material parameters it
could be as high as 100\%. The selection rules for spin injection
from allowed-allowed transitions are presented, and interband
spin-orbit coupling is found to play an important role.
\end{abstract}


\maketitle

\section{Introduction\label{sec:Intro}}

The optical injection of spin-polarized electrons in
semiconductors, familiar since the 1980s
\cite{DP_OpticalOrientation}, is again attracting attention, due
in part to the potential utilization of a spin-polarized
electrical current in a technology called ``spintronics"
\cite{AwschalomSpintronicsBook,Zutic04}. It is well known that
linear absorption of circularly polarized light in a semiconductor
produces spin-polarized electrons in the conduction band
\cite{DP_OpticalOrientation}. This occurs as a result of the
entanglement of electron spin and motion caused by the spin-orbit
coupling in the material; in the absence of spin-orbit coupling,
there would be no net spin polarization of the excited carriers.
For many common semiconductors, the highest valence states are in
the degenerate heavy and light hole bands at the $\Gamma$ point.
Consequently, the highest degree of spin polarization that can be
achieved is 50\%. Such a situation occurs when the photon energy
exceeds the band gap, but is not large enough to excite carriers
out of the split-off band. This can be understood from selection
rules that result from the symmetry of the states at the $\Gamma$
point \cite{DP_OpticalOrientation}.

One way to increase the spin polarization of the injected
electrons is to use materials where the degeneracy between heavy
and light hole bands is removed by strain and/or quantum
confinement, so that one can excite carriers only from one band.
From the symmetry of the states, one then expects 100\% spin
polarization. And indeed both theory \cite{DP_OpticalOrientation}
and experiments \cite{Maruyama91, Omori91} have shown a
significant enhancement of the degree of spin polarization. The
spin polarization could also be increased by using compounds with
crystal structures where there is no valence band degeneracy
\cite{Ciccaci87,Wei94,Janotti02}.

Spin injection can also arise from two-photon absorption. For
certain applications this may have advantages over one-photon spin
injection due to a much longer absorption depth, which allows spin
excitation throughout the volume of a bulk sample. Two-photon spin
injection has been investigated in lead chacogenides (PbTe, PbSe,
and PbS), which are cubic, and have direct fundamental band gaps
at the $L$ points \cite{Ivchenko73}. High degrees of spin
polarization in these materials have been predicted
\cite{Ivchenko73}, but not observed
\cite{Danishevskii78,Danishevskii87,Bresler87,Bresler88}. Our
focus in this paper is on semiconductors that have a direct
fundamental band gap at the $\Gamma$ point, such as GaAs. Based on
arguments involving the conservation of angular momentum, it was
recently suggested that 100\% spin polarization could be achieved
in unstrained bulk GaAs from two-photon absorption
\cite{Matsuyama01}. Earlier theoretical calculations, however,
predict a two-photon spin polarization of no more than 64\% for
this class of cubic semiconductors
\cite{Danishevskii72,Ivchenko73,Arifzhanov75,BhatSipe00}.

In this paper, we report results of time-resolved pump-probe
experiments that show the degrees of spin polarization from one-
or two-photon absorption are in fact comparable for GaAs. We also
discuss in detail the various effects that can complicate the
direct experimental comparison of the spin polarization obtained
by one- and two-photon excitation. We present microscopic
calculations of two-photon spin injection that go beyond the
spherical approximation made by earlier calculations. We show how
the simple argument based on conservation of angular momentum
breaks down, and examine the transitions that give rise to the
partial spin polarization. The calculated one- and two-photon
degrees of spin polarization are not equal for all materials, and
we find that, in fact, two-photon spin injection can be fully
polarized, but only from transitions that do \emph{not} conserve
angular momentum.

\begin{figure}
\includegraphics[width=3.1in]{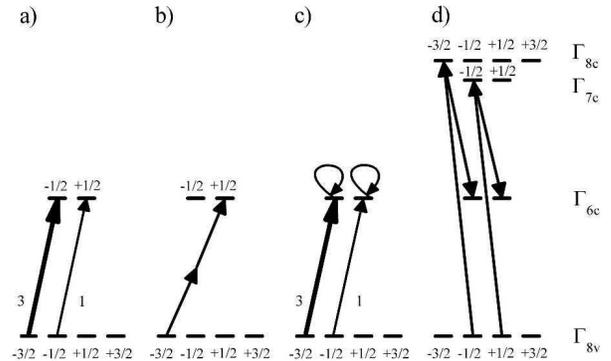}
\caption{Optical transitions in a bulk zincblende semiconductor
for circularly polarized light $\sigma^{+}$ allowed by the
selection rules: (a) for one-photon absorption, (b) for two-photon
absorption as suggested by Matsuyama \textit{et al}.
\cite{Matsuyama01}, (c) two-photon allowed-forbidden transitions
with a conduction band as an example of an intermediate state, and
(d) two-photon allowed-allowed transitions for vanishing interband
spin-orbit coupling and light incident along a $\langle001\rangle$
direction. The quantum number $m_{j}$ for the projection of total
angular momentum on the light propagation direction of all states
involved is indicated in the figures. The thickness of arrows and
adjacent number in (a) and (c) express the relative transition
probabilities.} \label{Fig_transitions}
\end{figure}

Optical transitions near the $\Gamma $ point can be summarized
with sketches such as those in Fig.\ \ref{Fig_transitions}. The
symmetry of the states at the $\Gamma $ point of a crystal with
zincblende symmetry is as follows. The conduction band
($\Gamma_{6c}$) is $s$-like with two degenerate spin states, while
the valence bands are $p$-like. The $p$-like orbitals are coupled
to the electron spin to form four states (the heavy and light hole
bands, $\Gamma_{8v}$) with total angular momentum $3/2$ and two
states at lower energy (the split-off band, $\Gamma_{7v}$) with
total angular momentum $1/2$ \cite{YuCardonaChapter2}. The levels
corresponding to the split-off band are not shown in Fig.\
\ref{Fig_transitions}. The selection rules for the transitions
between these states are the same as for the states of a
spherically symmetric system \cite{DP_OpticalOrientation}. Thus
they can be understood using angular momentum arguments. By
applying the familiar selection rule that one-photon absorption of
circularly polarized light with positive helicity ($\sigma^{+}$)
must change the projection of total angular momentum by $+1$, one
sees that only the two transitions shown in Fig.\
\ref{Fig_transitions}a are allowed. An examination of
Clebsch-Gordan coefficients reveals that the transition from the
$m_{j}=-3/2$ state of the valence band to the $m_{j}=-1/2$ state
of the conduction band is three times as probable as the
transition from the $m_{j}=-1/2$ state of the valence band to the
$m_{j}=+1/2$ of the conduction band. Thus, near the band edge, one
expects a value of $50\%$ for the degree of electron spin
polarization
\begin{equation}
P\equiv \frac{N_{\downarrow}-N_{\uparrow}}
 {N_{\downarrow}+N_{\uparrow}},
\label{DSPdefined}
\end{equation}
where $N_{\downarrow}$ ($N_{\uparrow}$ ) is the concentration of
electrons with spin down (up).

The idea of angular momentum conservation was applied to
two-photon absorption by Matsuyama \textit{et al}
\cite{Matsuyama01}. They argue that because the total angular
momentum of the two right circularly polarized photons is $+2$,
only the transition from $m_{j}=-3/2$ to $m_{j}=+1/2$ is allowed
(see also Fig.\ \ref{Fig_transitions}b). Therefore, they suggest
that even in a bulk semiconductor with degenerate valence bands
the resulting electron spin polarization should be 100\%, and
indeed with an opposite sign with respect to one-photon spin
injection.

On the other hand, the degree of spin polarization due to
two-photon spin injection has been calculated several times
\cite{Danishevskii72,Ivchenko73,Arifzhanov75,BhatSipe00} using the
8 band Kane model \cite{Kane57}. Ivchenko calculated the degree of
spin polarization in the limit of large spin-orbit splitting
\cite{Danishevskii72,Ivchenko73}. Arifzhanov and Ivchenko improved
the calculation by allowing the split-off band to act as an
intermediate state; they gave the degree of spin polarization at
the band edge as a function of $E_{g}/\Delta$, where $E_{g}$ is
the band gap energy and $\Delta$ is the spin-orbit splitting
\cite{Arifzhanov75}. For GaAs, one infers a 51\% degree of spin
polarization from their results. Note that, in contrast to
one-photon spin injection, the degree of spin polarization of
two-photon spin injection near the band edge depends not only on
the symmetry of the states, but also on various material
parameters. When only two-band transitions are included, a very
simple expression for the two-photon degree of spin polarization
has been given in terms of the conduction and valence band
effective masses \cite{BhatSipe00}, from which one infers for GaAs
a spin polarization of 48\%.

In most III-V semiconductors, the next higher conduction bands are
$p$-like ($\Gamma_{7c}$ and $\Gamma_{8c}$)
\cite{YuCardonaChapter2}. The role that these higher bands play in
two-photon spin injection has not previously been investigated. It
is known that $\mathbf{k}\cdot \mathbf{p}$ mixing with these bands
is responsible for cubic anisotropy of two-photon absorption
\cite{Dvorak94,HW94}. The higher conduction bands can also act as
intermediate states in the two-photon amplitude. Such transitions
are qualitatively different than transitions within the set of
bands nearest to the fundamental band gap
\cite{vanderZiel77,Catalano86,LeeFan74,Bredikhin73}.

In section \ref{sec:calc} we review the symmetry of two-photon
spin injection and present our calculation including the higher
conduction bands. In contrast to previous calculations of
two-photon spin injection
\cite{Danishevskii72,Ivchenko73,Arifzhanov75,BhatSipe00}, our
calculation is not perturbative in $k$. In section
\ref{sec:experiment} we present the experimental comparison of
one- and two-photon spin injection. In section
\ref{sec:discussion} we discuss the transitions responsible for
the degree of spin polarization in two-photon absorption (Figs.\
\ref{Fig_transitions} c and d). In an appendix we derive
expressions for the degree of two-photon spin injection due to
so-called `allowed-allowed' transitions.

\section{Calculation of two-photon spin injection\label{sec:calc}}

For an electric field of the form $\mathbf{E}\left( t\right)
=\mathbf{E}_{\omega}\exp (-i\omega t)+c.c.$ (we sometimes write
$\mathbf{E}_{\omega}=E_{\omega} \mathbf{\hat{e}}_{\omega}$), the
two-photon spin injection rate can be written phenomenologically
as $\dot{S}^{i}=\zeta^{ijklm}E_{\omega }^{j}E_{\omega
}^{k}E_{\omega }^{l*}E_{\omega }^{m*}$ where $\zeta^{ijklm}$ is a
fifth rank pseudotensor symmetric on exchange of indices $j$ and
$k$, and on exchange of indices $l$ and $m$; superscript lowercase
letters denote Cartesian components and repeated indices are to be
summed over \cite{Najmaie03}. For the point groups $T_{d}$ and
$O_{h}$, appropriate to most cubic semiconductors, a general fifth
rank pseudotensor has ten independent components. Forcing the
$j\leftrightarrow k$ and $l\leftrightarrow m$ symmetries, and the
condition for reality of $\dot{\mathbf{S}}$, $\zeta^{ilmjk} =
(\zeta^{ijklm})^{*}$, leaves three independent real components.

We define $\zeta_{2A} \equiv -i\zeta^{abccc}$ and $\zeta_{2B}
\equiv \mathrm{Im}\zeta^{aabac}$, where the indices $a$, $b$, and
$c$ denote components along the standard cubic axes $[100]$,
$[010]$, and $[001]$. Then the three independent real components
are $\mathrm{Re}\zeta_{2A}$, $\mathrm{Im}\zeta_{2A}$, and
$\zeta_{2B}$. In the standard cubic basis, the non-zero components
of $\zeta $ are
\begin{eqnarray*}
\zeta^{abbbc} &=& \zeta^{abccc} = \zeta^{bacaa} = i\zeta_{2A} \\
\zeta^{bccac} &=& \zeta^{caaab} = \zeta^{cabbb} = i\zeta_{2A} \\
\zeta^{abcbb} &=& \zeta^{accbc} = \zeta^{baaac} = -i\zeta_{2A}^{*} \\
\zeta^{baccc} &=& \zeta^{cabaa} = \zeta^{cbbab} = -i\zeta_{2A}^{*} \\
\zeta^{aabac} &=& \zeta^{bbcab} = \zeta^{cacbc} = i\zeta_{2B} \\
\zeta^{aacab} &=& \zeta^{babbc} = \zeta^{cbcac} = -i\zeta_{2B},
\end{eqnarray*}
as well as those generated by exchanging $j\leftrightarrow k$
and/or $l\leftrightarrow m$, for a total of 48 components.

The point group symmetry allows spin injection for linearly
polarized light, associated with $\mathrm{Im}\zeta_{2A}$. However,
from a microscopic expression for $\zeta $ in the independent
particle picture [see Eq.\ (\ref{Sdot_doublyDengenerate}),
(\ref{Sdot_no_spinsplitting}), or (\ref{Sdot_w_spinsplitting})
below], one can show that $\zeta$ must be purely imaginary due to
the time reversal properties of the Bloch states. One might expect
deviations from the independent particle picture within an exciton
binding energy of the band edge \cite{Mahan68,LeeFan74}. In what
follows, we assume the independent particle picture is valid,
which leaves the two-photon spin injection specified in terms of
two real parameters $\zeta_{2A}$ and $\zeta_{2B}$.

The component of the spin injection rate along one of the cubic
axes can be written compactly (with no summation convention) as
\cite{Ivchenko73}
\begin{equation}
    \dot{S}^{i}=2i \left(\mathbf{E}_{\omega}\times \mathbf{E}_{\omega}^{*}\right)^{i}
    \left( \zeta_{2A} \left|\mathbf{E}_{\omega}\right|^{2} +
    \left( 2\zeta_{2B} - \zeta_{2A}\right)
    \left| E_{\omega }^{i}\right|^{2}\right).\label{sDotPhenomCubic}
\end{equation}
If the material were isotropic, the spin injection rate could be
described by only one real parameter; $\zeta_{2A}=2\zeta_{2B}$ and
the second term in Eq.\ (\ref{sDotPhenomCubic}) would be zero.

The cubic anisotropy means that the two-photon spin injection from
circularly polarized light depends on the angle of incidence of
the light relative to the cubic axes. For circularly polarized
light incident along $\mathbf{\hat{n}}$ specified by polar angles
$\theta $ and $\phi $ relative to the cubic axes,
\begin{equation} \dot{\mathbf{S}}\cdot
\mathbf{\hat{n}}=\mp 2\zeta_{2A}|E_{\omega}|^{4}\left(
1+\frac{2\zeta_{2B}-\zeta_{2A}}{4\zeta_{2A}}f\left( \theta ,\phi
\right) \right) , \label{SdotMacroCirc}
\end{equation}
where $f\left( \theta ,\phi \right) =\sin^{2}\left( 2\theta
\right) +\sin^{4}\left( \theta \right) \sin^{2}\left( 2\phi
\right) $. The upper (lower) sign is for right (left) circular
polarization. The analogous equation for two-photon absorption is
given by Hutchings and Wherrett \cite{HW94}. Equation
(\ref{SdotMacroCirc}) has extrema for light incident along
$\left\langle 001\right\rangle $ and $\left\langle
111\right\rangle $ directions. Due to the cubic anisotropy, the
net injected spin is not always parallel to $\mathbf{\hat{n}}$,
although it is when $\mathbf{\hat{n}}$ is along $\left\langle
001\right\rangle $ or $\left\langle 111\right\rangle $. In
particular, for light along a $\left\langle 001\right\rangle $
direction, $| \dot{\mathbf{S}}| =2\zeta_{2A}|E_{\omega }|^{4}$,
while for light incident along a $\left\langle 111\right\rangle $
direction, $| \dot{\mathbf{S}}| =\left( 4/3\right) \left(
\zeta_{2A}+\zeta_{2B}\right) |E_{\omega }|^{4}$.

\subsection{Microscopic calculation}\label{subsec:micro}

We calculate the photoinjection rate of net electron spin density,
$\dot{\mathbf{S}}$, using second order perturbation theory with
the light treated classically in the long wavelength limit. We
ignore interactions amongst the electrons, and between electrons
and phonons. We take the photon energy to be below the band gap,
and twice the photon energy to be above the band gap. We neglect
any spin polarization of the holes, since their spin relaxation
times are typically very short \cite{HiltonTang}. In the Fermi's
Golden Rule (FGR) limit, the photoinjection rate is
time-independent.

Expressions for the two-photon spin injection rate under these
assumptions have been given before
\cite{Ivchenko73,Arifzhanov75,BhatSipe00,Najmaie03}. However, all
previous calculations used semiconductor models in which all bands
are doubly degenerate. In such a case, one finds that
\begin{equation}
\dot{\mathbf{S}}=\frac{2\pi }{L^{3}}\sum_{c,c^{\prime
},v,\mathbf{k} }^{\prime }\left\langle c\mathbf{k}\right|
\mathbf{\hat{S}}\left| c^{\prime }\mathbf{k} \right\rangle \Omega
_{c,v,\mathbf{k}}^{\left( 2\right) *}\Omega_{c^{\prime
},v,\mathbf{k}}^{\left( 2\right) }\delta \left[ 2\omega -\omega
_{cv}\left( \mathbf{k}\right) \right] ,
\label{Sdot_doublyDengenerate}
\end{equation}
where $\left| n\mathbf{k}\right\rangle $ is a Bloch state with
energy $\hbar \omega_{n}( \mathbf{k}) $, $L^{3}$ is a
normalization volume, $ \mathbf{\hat{S}}$ is the spin operator,
$\omega_{nm}( \mathbf{k} ) \equiv \omega_{n}( \mathbf{k} )
-\omega_{m} ( \mathbf{k} ) $, the prime on the summation indicates
a restriction to pairs $(c,c^{\prime})$ for which
$\omega_{cc^{\prime}}=0$, and $\Omega_{c,v,\mathbf{k}}^{(2)}$ is
the two-photon amplitude
\begin{equation}
\Omega_{c,v,\mathbf{k}}^{(2)} = \left( \frac{e}{\hbar
\omega}\right)^{2} \sum_{n}\frac{\left( \mathbf{E}_{\omega }\cdot
\mathbf{v}_{c,n}\left( \mathbf{k}\right) \right) \left(
\mathbf{E}_{\omega }\cdot \mathbf{v}_{n,v}\left( \mathbf{k}\right)
\right) }{\omega_{nv} -\omega \left( \mathbf{k}\right) },
\label{TPamplitude}
\end{equation}
with $\mathbf{v}_{n,m}\left( \mathbf{k}\right) \equiv \left\langle
n\mathbf{k}\right| \mathbf{\hat{v}}\left| m\mathbf{k}\right\rangle
$ where $\mathbf{\hat{v}}$ is the velocity operator. It is well
known, however, that in real crystals of zincblende symmetry the
spin degeneracy is removed \cite{Dresselhaus,PikusReview88},
albeit with a small energy splitting. Since we are using a model
that accounts for this spin-splitting \cite{PZ96}, we must
generalize the earlier microscopic expressions. Such a
generalization was recently discussed for one-photon spin
injection \cite{LinearPSC04}.

If the spin-split bands are well separated, FGR gives
\begin{equation}
\dot{\mathbf{S}}=\frac{2\pi }{L^{3}}\sum_{c,v,\mathbf{k}}\left\langle c%
\mathbf{k}\right| \mathbf{\hat{S}}\left| c\mathbf{k}\right\rangle
\left| \Omega_{c,v,\mathbf{k}}^{\left( 2\right) }\right|
^{2}\delta \left[ 2\omega -\omega_{cv}\left( \mathbf{k}\right)
\right] . \label{Sdot_no_spinsplitting}
\end{equation}
However, in GaAs the splitting is at most a few meV for conduction
states within 500\,meV of the band edge \cite{CCF88}. Since this
is comparable to the broadening that one would calculate from the
scattering time of the states (and also to the laser bandwidth for
experiments with pulses shorter than 100\,fs), spin-split pairs of
bands should be treated as quasi-degenerate in FGR\@. Thus in
place of Eq.\ (\ref{Sdot_doublyDengenerate}) or
(\ref{Sdot_no_spinsplitting}) we use,
\begin{equation}
\begin{split}
\dot{\mathbf{S}} =&\frac{2\pi }{L^{3}}\sum_{c,c^{\prime
},v,\mathbf{k} }^{\prime }\left\langle c\mathbf{k}\right|
\mathbf{\hat{S}}\left| c^{\prime } \mathbf{k}\right\rangle \Omega
_{c,v,\mathbf{k}}^{\left( 2\right) *}\Omega_{c^{\prime
},v,\mathbf{k}}^{\left( 2\right) }
\\
&\times \frac{1}{2}\left\{ \delta \left[ 2\omega -\omega
_{cv}\left( \mathbf{k}\right) \right] +\delta \left[ 2\omega
-\omega_{c^{\prime }v}\left( \mathbf{k}\right) \right] \right\} ,
\end{split}
\label{Sdot_w_spinsplitting}
\end{equation}
where the prime on the summation indicates a restriction to pairs
$\left( c,c^{\prime }\right) $ for which either $c^{\prime }=c$,
or $c$ and $c^{\prime }$ are a quasi-degenerate pair. The
coherence between quasi-degenerate bands is optically excited and
grows with their populations, as is the case with simpler band
models that neglect spin splitting
\cite{Danishevskii72,Ivchenko73,Arifzhanov75,BhatSipe00,Najmaie03}.
Using the time reversal properties of the Bloch functions, the
expression for $\zeta^{ijklm}$ that follows from
(\ref{Sdot_w_spinsplitting}) can be simplified to give
\begin{equation}
\begin{split}
\zeta^{ijklm}=&i\left( \frac{e}{\hbar \omega }\right)
^{4}\frac{2\pi }{L^{3}} \sum_{c,c^{\prime },v,\mathbf{k}}^{\prime
}\sum_{n,n^{\prime }} \delta \left[ 2\omega -\omega _{cv}\left(
\mathbf{k}\right) \right]
\\
&\times \mathrm{Im} \left[ \frac{\left\langle c\mathbf{k}\right|
\hat{S}^{i}\left| c^{\prime}\mathbf{k}\right\rangle \left(
V^{jklm}-V^{lmjk}\right) /2}{\left( \omega _{nv}\left(
\mathbf{k}\right) -\omega \right) \left( \omega _{n^{\prime
}v}\left( \mathbf{k}\right) -\omega \right) }\right] ,
\end{split}
\label{eqn_zeta_micro}
\end{equation}
where
\begin{equation*}
V^{jklm}\equiv\left\{ \mathbf{v}_{c^{\prime },n^{\prime }}\left( \mathbf{k}%
\right) ,\mathbf{v}_{n^{\prime },v}\left( \mathbf{k}\right)
\right\}
^{jk}\left\{ \mathbf{v}_{c,n}^{*}\left( \mathbf{k}\right) ,\mathbf{v}%
_{n,v}^{*}\left( \mathbf{k}\right) \right\} ^{lm} ,
\end{equation*}
and $\left\{ \mathbf{v}_{1},\mathbf{v}_{2}\right\} ^{ij}\equiv(
v_{1}^{i}v_{2}^{j}+v_{1}^{j}v_{2}^{i}) /2$.

The photoinjection rate for the density of electron-hole pairs is
\begin{equation}
\dot{N}=\frac{2\pi }{L^{3}}\sum_{c,v,\mathbf{k}}\left|
\Omega_{c,v,\mathbf{k}}^{\left( 2\right) }\right|^{2}\delta \left[
2\omega -\omega_{cv}\left( \mathbf{k}\right) \right] .
\label{Ndot}
\end{equation}
From Eqs.\ (\ref{Sdot_w_spinsplitting}) and (\ref{Ndot}), the
degree of spin polarization, $P$, can be calculated, since
\begin{equation}
P=-\frac{2}{\hbar }\frac{\dot{\mathbf{S}} \cdot \mathbf{\hat{n}}
}{\dot{N}}. \label{DSP_micro}
\end{equation}
The sign of $P$ is chosen so that a positive $P$ corresponds to an
excess of electrons with spin down, i.e. spin opposite the photon
angular momentum.

To evaluate the degree of spin polarization, we use a
$\mathbf{k}\cdot \mathbf{p}$ model that diagonalizes the
one-electron Hamiltonian (including spin-orbit coupling) within a
basis set of 14 $\Gamma $ point states, and includes important
remote band effects \cite{PZ96}. Fourteen band models (also called
five-level models) have been used to calculate bandstructures
\cite{Rossler84,PZ90,MayerRossler91,MayerRossler93b}, as well as
linear \cite{MayerRossler93,LinearPSC04} and non-linear
\cite{HW94,Hutchings95,Hutchings97} optical properties of GaAs and
other semiconductors. Winkler has given a recent review of 14 band
models \cite{WinklerBook}. The 14 states (counting one for each
spin) comprise six valence band states (the split-off, heavy, and
light hole bands), and eight conduction band states (the two
lowest, which are $s$-like, and the six next-lowest, which are
$p$-like). The states are given in more detail in Appendix
\ref{sec:Notation}, and except for the split-off hole states, they
are shown in Fig.\ \ref{Fig_transitions}d.

The model contains 13 parameters chosen to fit low-temperature
experimental data. Of the two parameter sets discussed by Pfeffer
and Zawadzki for GaAs, we use the one corresponding to $\alpha
=0.085$ that they find gives better results \cite{PZ96}. For InP,
GaSb, and InSb, we use parameters from Cardona, Christensen and
Fasal \cite{CCF88}. The parameters are listed in Table
\ref{Table:parameters} and the notation is described in Appendix
\ref{sec:Notation}. For cubic ZnSe, we use the parameters given by
Mayer and Rossler \cite{MayerRossler93b} and a calculated value of
$C_{k}$ \cite{CCF88}; we use $\Delta^{-}=-0.238$\,eV to give a
$k^{3}$ conduction band spin-splitting that matches the \textit{ab
initio} calculation of Cardona, Christensen and Fasal
\cite{CCF88}. There is more uncertainty in the parameters for ZnSe
than in those for the other materials \cite{MayerRossler93b}, but
we include it as an example of a semiconductor with a larger band
gap.

\begin{table}
\caption{Model parameters. \label{Table:parameters}}
\begin{ruledtabular}
\begin{tabular}{c|ddddd}
    &\multicolumn{1}{c}{GaAs}  & \multicolumn{1}{c}{InP}
     & \multicolumn{1}{c}{GaSb} & \multicolumn{1}{c}{InSb} &  \multicolumn{1}{c}{ZnSe}\\
\hline
$E_{g}$ (eV) & 1.519 & 1.424 & 0.813 & 0.235 & 2.820\\
$\Delta_{0}$ (eV) & 0.341 & 0.108 & 0.75 & 0.803 & 0.403\\
$E_{0}^{\prime}$ (eV) & 4.488 & 4.6 & 3.3 & 3.39 & 7.330\\
$\Delta_{0}^{\prime}$ (eV) & 0.171 & 0.50 & 0.33 & 0.39 & 0.090\\
$\Delta^{-}$ (eV) & -0.061 & 0.22 & -0.28 & -0.244 & -0.238\\
$P_{0}$ (eV\AA) & 10.30 & 8.65 & 9.50 & 9.51 & 10.628\\
$Q$ (eV\AA) & 7.70 & 7.24 & 8.12 & 8.22 & 9.845\\
$P_{0}^{\prime}$ (eV\AA) & 3.00 & 4.30 & 3.33 & 3.17 & 9.165\\
$\gamma_{1L}$ & 7.797 & 5.05 & 13.2 & 40.1 & 4.30\\
$\gamma_{2L}$ & 2.458 & 1.6 & 4.4 & 18.1 & 1.14\\
$\gamma_{3L}$ & 3.299 & 1.73 & 5.7 & 19.2 & 1.84\\
$F$ & -1.055 & 0 & 0 & 0 & 0\\
$C_{k}$ (meV\AA) & -3.4 & -14 & 0.43 & -9.2 & -14
\end{tabular}
\end{ruledtabular}
\end{table}

Note that although remote band terms are included in the $14\times
14$ Hamiltonian, we have neglected the remote band contributions
to the velocity operator. The effect of these contributions on
one-photon absorption was discussed by Enders \textit{et
al}.\cite{Enders95} Removing the remote band terms from our
Hamiltonian changes $P$ for GaAs by at most $2\%$. Thus we feel
justified in our neglect of the remote band contributions to the
velocity operator. Another contribution to the velocity operator
$\mathbf{\hat{v}}$, the anomalous velocity term, $\hbar \left(
\bm{\sigma }\times \bm{\nabla } V\right) /\left(
4m^{2}c^{2}\right) $ should be included if $k$-dependent
spin-orbit coupling is included in the $14\times 14$ Hamiltonian.
For the results reported in the following section, we have
neglected $k$-dependent spin-orbit coupling. To test whether this
neglect is justified, we have repeated the calculation for GaAs
including such coupling only between valence and lowest conduction
bands and the associated anomalous velocity; the coupling is
parameterized by $C_{0}=0.16$\,eV{\AA} \cite{Ostromek96} (note
that $C_{0}$ is distinct from the $k$-linear term $C_{k}$). It
decreases the two-photon $P$ by $\approx 2\%$ for excess energies
between 0.1 and 200\,meV\@. The decrease increases for larger
excess energy, reaching $\approx 5\%$ for an excess energy of
500\,meV\@.

Our two-photon spin injection calculation is similar to the
two-photon absorption calculation of Hutchings and Wherrett
\cite{HW94}. We can reproduce their results by removing remote
band effects, which they did not include.

\subsection{Calculation results\label{subsec:calcresults}}

\begin{figure}
\includegraphics[width=3.1in]{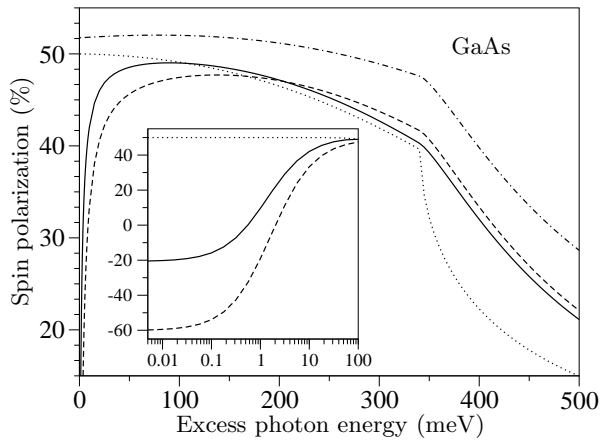}
\caption{Calculated degree of electron spin polarization $P$ in
GaAs. The solid (dashed) line is for two-photon excitation with
light incident along a $\left\langle 001\right\rangle $
($\left\langle 111\right\rangle $) direction and the dotted line
is for one-photon excitation. The dash-dotted line is for
two-photon excitation calculated with an 8 band ($\Gamma_{7v}$,
$\Gamma_{8v}$, and $\Gamma_{6c}$) spherical model. The inset shows
$P$ close to the band edge. The sign of $P$ is given by Eq.\
(\ref{DSP_micro}).} \label{Fig_DSP_Calc}
\end{figure}

\begin{figure}
\includegraphics[width=3.1in]{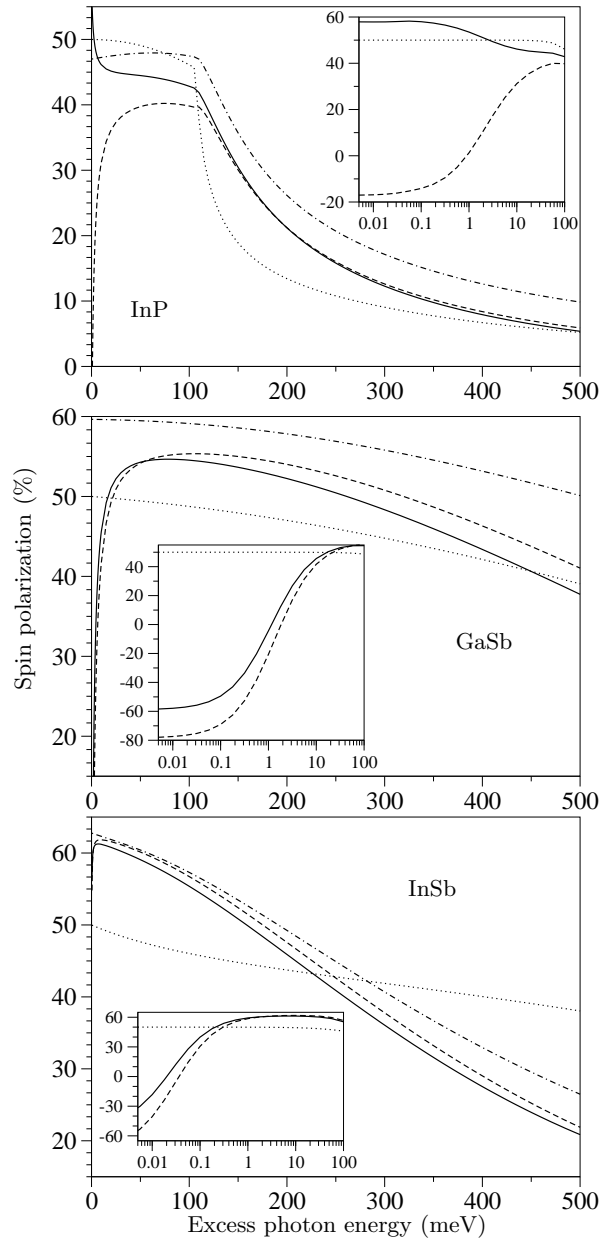}
\caption{As Fig.\ \ref{Fig_DSP_Calc}, but for InP, GaSb, and
InSb.} \label{Fig_InP_GaSb_InSb}
\end{figure}

\begin{figure}
\includegraphics[width=3.1in]{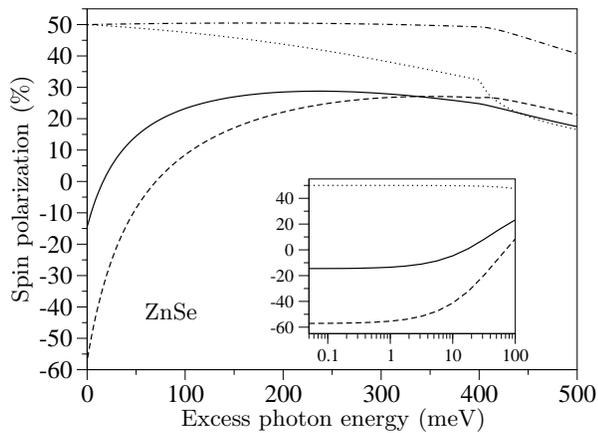}
\caption{As Fig.\ \ref{Fig_DSP_Calc}, but for cubic ZnSe.}
\label{Fig_ZnSe}
\end{figure}

The calculated degrees of electron spin polarization, $P$, are
shown for GaAs, InP, GaSb, InSb, and cubic ZnSe in Figs.\
\ref{Fig_DSP_Calc}--\ref{Fig_ZnSe} as a function of excess photon
energy, $2\hbar \omega -E_{g}$, where $E_{g}$ is the fundamental
band gap. We also show, for comparison, the degree of electron
spin polarization due to one-photon absorption \cite{LinearPSC04}.
For each semiconductor, the one-photon degree of spin polarization
is 50\% at the band edge as expected from the $\Gamma $ point
selection rules.

In GaAs, so long as the excess photon energy is less than the
split-off energy ($341$\,meV) and greater than about $50$\,meV,
there is a near equality of one- and two-photon $P$'s.

Close to the band edge however, there is a feature of the
two-photon $P$ that has not previously been identified; it is seen
more clearly in the insets of Figs.\
\ref{Fig_DSP_Calc}--\ref{Fig_ZnSe}. The values of the two-photon
$P$ at the band edge for each material are listed in Table
\ref{Table:bandedgeDSP}. We discuss this feature further in Sec.
\ref{sec:discussion}, but we note here that it does not appear in
a spherical approximation. To show this, we have calculated the
two-photon $P$ with the $8 \times 8$ Kane model that includes only
the valence bands and the $\Gamma_{6c}$ conduction bands, and with
$\gamma_{3}$ set equal to $\gamma_{2}$ and $C_{k}=0$; the result,
which is independent of crystal orientation, is shown in the
dash-dotted line in Figs.\ \ref{Fig_DSP_Calc}--\ref{Fig_ZnSe}.

\begin{table}
\caption{Calculated band-edge two-photon $P$.
\label{Table:bandedgeDSP}}
\begin{ruledtabular}
\begin{tabular}{c|ddddd}
     &\multicolumn{1}{c}{GaAs}  & \multicolumn{1}{c}{InP}
     & \multicolumn{1}{c}{GaSb} & \multicolumn{1}{c}{InSb}
     & \multicolumn{1}{c}{ZnSe}\\
      \hline
$[001]$ & -20.5\% & 58.7\% & -58.9\% & -49.3\% & -14.5\%\\
$[111]$ & -60.0\% & -16.6\%& -78.4\% & -73.3\% & -57.1\%\\
\end{tabular}
\end{ruledtabular}
\end{table}

Both one- and two-photon $P$'s decrease as the excess photon
energy is increased. This is due to band mixing away from the
$\Gamma $ point, which changes the selection rules. At excess
photon energies above the split-off energy, the one-photon $P$
decreases due to transitions from the split-off valence band
\cite{DP_OpticalOrientation}. The two-photon $P$ also decreases
due to these transitions, but less so.

The possibility of cubic anisotropy in two-photon spin injection
was first pointed out by Ivchenko \cite{Ivchenko73}, although it
has not been calculated until now. Cubic anisotropy in two-photon
absorption, on the other hand, has been calculated by Hutchings
and Wherrett \cite{HW94}. They found that near the band edge
two-photon absorption of circularly polarized light in GaAs should
be about 10\% greater for light incident along $\left[ 111\right]
$ compared to along $\left[ 001\right] $ \cite{HW94}. The results
of our calculation for GaAs indicate that two-photon spin
injection varies with crystal orientation by a similar amount.
Hence the degree of spin polarization shown in Fig.\
\ref{Fig_DSP_Calc}, which is the ratio of the two, varies with
crystal orientation by only a few percent for most photon energies
in the range we investigated. This is not the case, however, for
excess photon energies very close to the band gap, as can be seen
in the inset of Fig.\ \ref{Fig_DSP_Calc}. The cubic anisotropy is
more substantial for ZnSe and InP\@.

\section{Experimental comparison\label{sec:experiment}}

To experimentally measure the degree of electron spin polarization
we performed a polarization-resolved pump-probe experiment, where
the transmission of the probe pulses is measured as a function of
the delay between $\approx 150$\,fs circularly polarized pump and
probe pulses. Specifically, we measure the differential
transmission $\Delta T/T=(T_{E}-T_{0})/T_{0}$, where $T_{E}$
($T_{0}$) is the transmission with (without) the pump. If the
absorbance change induced by the pump is small (i.e., $\Delta
\alpha l \ll 1$, where $l$ is the sample thickness and $\Delta
\alpha = \alpha_{E} - \alpha_{0}$ is the difference between the
absorption coefficient with and without the pump, respectively),
the differential transmission will be proportional to $-\Delta
\alpha l$. Furthermore, if this weak absorption change is caused
by phase-space filling associated with a thermalized nondegenerate
distribution of carriers, then the differential transmission will
be proportional to the carrier density ($\Delta T/T \propto N$).
These conditions are usually satisfied after about 0.5 ps for thin
samples, low carrier densities and relatively high temperatures.
Finally, if the holes do not contribute significantly to the
phase-space filling, then the degree of polarization can be
experimentally determined by measuring the differential
transmission for pump and probe pulses having the same $(\Delta
T/T)^{++}$ and opposite $(\Delta T/T)^{+-}$ circular
polarizations. For probe pulses near the band edge, $(\Delta
T/T)^{++}\propto 3N_{\downarrow}+N_{\uparrow}$ and $(\Delta
T/T)^{+-}\propto 3N_{\uparrow }+N_{\downarrow}$, as a result of
the same selection rules described above in Sec. \ref{sec:Intro}.
Defining,
\begin{equation}
P_{\textrm{exp}}\equiv 2\frac{\left( \Delta T/T\right)^{++}-\left(
\Delta T/T\right)^{+-}}{\left( \Delta T/T\right)^{++}+\left(
\Delta T/T\right)^{+-}}, \label{PexpEqn}
\end{equation}
we then have $P_{\textrm{exp}}=P$. However, if these restricted
conditions are not met, then $P_{\textrm{exp}}$ may not directly
yield the degree of polarization. The degree of polarization and
the spin relaxation time may still be extracted, but other effects
may have to be considered \cite{Yaser_above_bandgap}.

Polarization-resolved differential transmission measurements were
performed using pulses from an optical parametric amplifier (OPA)
\cite{Reed96} pumped by a regeneratively amplified Ti:sapphire
laser operating at 250\,kHz. The laser system was tuned to produce
$\approx 150$\,fs pulses at 1550\,nm (signal) and 1650\,nm
(idler). Two beta barium borate (BBO) crystals were used to
generate 775\,nm pulses from the signal beam and 825\,nm pulses
from the idler beam. The second-harmonic and fundamental pulses
were then separated using dichroic beamsplitters. Thus, we used
775\,nm pulses to excite the sample by one-photon absorption,
1550\,nm pulses to excite the sample by two-photon absorption and
825\,nm pulses to probe the transmission of the sample.

We used a semi-insulating (impurity level less than
$10^{15}$\,cm$^{-3}$), 1\,$\mu$m thick sample of $[001]$-grown
bulk GaAs that was van der Waals bonded to the glass substrate.
The experiments were performed at a temperature of 80\,K\@.
Consequently, the probe beam was resonant with the band gap energy
$E_{g}$, and the pump beams had an excess energy ($2\hbar \omega
-E_{g}$) of 90\,meV, which is considerably less than the
spin-orbit splitting energy of 341\,meV\@. The peak irradiances of
pump pulses were $\approx 2.3$\,GW/cm$^{2}$ (fluence $\approx
320$\,mJ/cm$^{2}$) for two-photon and $\approx 11$\,MW/cm$^{2}$
(fluence $\approx 1.2$\,mJ/cm$^{2}$) for one-photon absorption,
exciting in both cases a carrier density of $\approx 6\times
10^{16}$\,cm$^{-3}$. Probe pulses were 10 times weaker than pump
pulses for one-photon excitation.

The design and implementation of the experiments involved a
substantial effort to remove all possible experimental artifacts
that could influence direct comparison of the results obtained for
optical pumping by one- and two-photon absorption. First, by using
a pump-probe technique, we directly measured the degree of spin
polarization of electrons at fixed times after the generation
process. Thus our data are more credible than that of experiments
using time-integrated methods, where the measured degree of spin
polarization has to be corrected using the ratio of the lifetime
and the spin relaxation time of electrons
\cite{Danishevskii72,Matsuyama01}. Second, we used a relatively
thin layer of GaAs. One- and two-photon absorption have
substantially different excitation depth profiles, the former
being considerably steeper than the latter. In order to directly
use Eq.\ (\ref{PexpEqn}) to determine the degree of spin
polarization, it is necessary to keep the sample thin enough to
ensure that $\Delta \alpha l \ll 1$ for each process; however, the
sample must be kept as thick as possible to maximize the magnitude
of $\Delta T/T$. We selected a 1\,$\mu$m thickness as a compromise
between these two tendencies. For the sample temperature and the
wavelengths used here, the transmissions $T$ of pump pulses for
the two-photon excitation were larger than 99\%. For pump pulses
for one-photon excitation $T\approx 30\%$, and for the probe
pulses $T \approx 70\%$ \cite{PalikHandbook_p439}. Third, by using
a probe wavelength different from the pump, we were able to
improve signal to noise by using a spectral filter to eliminate
scattered pump light. Fourth, we precisely characterized the
quality of the circular polarization of all optical beams used.
Due to the finite spectral bandwidth of the femtosecond optical
pulses and the quality of the quarter wave ($\lambda /4$) plates,
the optical beams were not 100\% circularly polarized, but instead
consisted of both $\sigma^{+}$ and $\sigma^{-}$. However, the
quality of the circular polarization of both pumps beams was
nearly the same. A nominal $\sigma^{+}$ polarization state of pump
pulses for one-photon excitation was 95\% $\sigma^{+}$ and 5\%
$\sigma^{-}$, while for two-photon excitation a nominal
$\sigma^{+}$ polarization state was 94\% $\sigma^{+}$ and 6\%
$\sigma^{-}$. As another check, we used a broadband quarter wave
plate to monitor the helicity of all beams. Finally, to avoid
problems due to possible sample inhomogeneity, we focused all
three beams (the pump beams for one- and two-photon excitation,
and the probe beam) to the same position on the sample using a
single achromatic lens. Their mutual spatial overlap was checked
using a pinhole.

\begin{figure}
\includegraphics[width=3.3in]{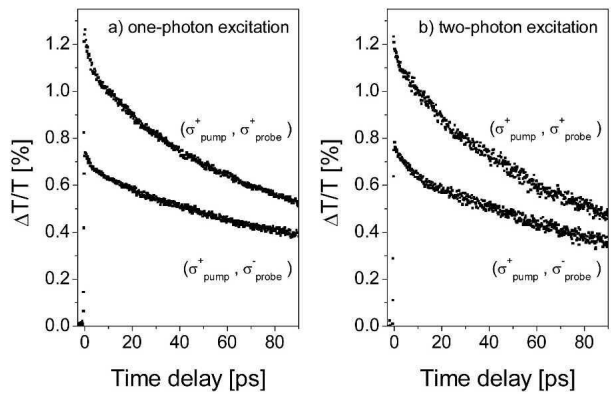}
\caption{The dynamics of differential transmission $\Delta T/T$
after excitation by circularly polarized pump pulses with the
excess energy of 90\,meV as measured using probe pulses with the
same (upper curve) and opposite (lower curve) circular
polarization for one-photon (a) and two-photon (b) excitation.}
\label{Fig_dT_T}
\end{figure}

\begin{figure}
\includegraphics[width=3.1in]{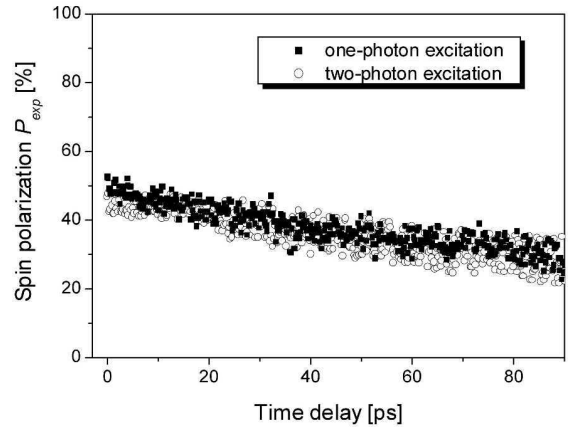}
\caption{The dynamics of the degree of spin polarization of
electrons $P_{\textrm{exp}}$ after one-photon (solid squares) and
two-photon (open circles) excitation as computed from data shown
in Fig.\ \ref{Fig_dT_T}.} \label{Fig_DSP_Exp}
\end{figure}

The results of the pump-probe experiment for one-photon excitation
by a $ \sigma^{+}$ pump are shown in Fig.\ \ref{Fig_dT_T}a. The
upper curve corresponds to probing with a $\sigma^{+}$ probe,
while the lower curve was measured with a $\sigma^{-}$ probe. The
difference between the different polarization conditions is caused
by spin-dependent phase-space filling as described above. The
resulting electron spin polarization $P$ as a function of time
delay is shown in Fig.\ \ref{Fig_DSP_Exp} (full squares).
The decay of $P$ is due to the randomization of the initial spin
polarization. From the data depicted in Fig.\ \ref{Fig_DSP_Exp} we
infer a time constant of $\approx 200$\,ps, which is
conventionally considered as half of the spin relaxation time. The
dominant spin relaxation mechanism is probably the precession
about anisotropic internal magnetic fields (the D'yakonov-Perel'
mechanism) \cite{DP71,KikkawaAwschalom98,Song02}. For two-photon
excitation by a $\sigma^{+}$ pump, the $\Delta T/T$ signal was the
same as in the case of one-photon excitation as illustrated in
Fig.\ \ref{Fig_dT_T}b. The resulting values of spin polarization
$P$ are shown in Fig.\ \ref{Fig_DSP_Exp} (open circles) as a
function of time delay. We can now directly compare the results
obtained for one- and two-photon excitation. First, note that
after excitation by $\sigma^{+}$ pump pulses $P$ has the
\textit{same sign} in both cases, in a clear contrast to the
predictions made by in Matsuyama \textit{et al}.
\cite{Matsuyama01} (see Fig.\ \ref{Fig_transitions}b). Second, the
initial values of $P$ for both one- and two-photon excitations
are, within the experimental error (including the nonideal
polarization state of optical pulses used), the same and equal to
the theoretical value of $49$\% expected for both at these photon
energies.

\section{Discussion\label{sec:discussion}}

The prediction of a 100\% degree of two-photon spin injection
mentioned in Section \ref{sec:Intro} uses arguments familiar from
spherically symmetric systems. At first it might seem incorrect to
even apply these in cubic systems. For the crystal Hamiltonian is
not rotationally invariant and thus does not conserve angular
momentum: The lattice is viewed as fixed and able to provide any
amount of torque. However, the deviation from spherical symmetry
is small in many cases, and hence angular momentum arguments
should have approximate validity. Stated more technically, since
$T_{d}$ is a subgroup of $O_{h}$, which is a subgroup of the full
rotation group, the Hamiltonian can be written as the sum of
spherical, cubic and tetrahedral parts with the latter two treated
as perturbations
\cite{Baldereschi70,Baldereschi71,Baldereschi73,Lipari70,WinklerBook}.
The 8 band Kane model (even including remote band effects but with
$\gamma_{2L}=\gamma_{3L}$ and $C_{k}=0$) is spherically symmetric
and is often used to describe many properties. It has been used,
in particular, for earlier calculations of one- and two-photon
spin injection
\cite{DP_OpticalOrientation,Danishevskii72,Ivchenko73,Arifzhanov75,BhatSipe00}.
In a spherical model however, the transitions depicted in Fig.\
\ref{Fig_transitions}b do not occur. By examining the possible
intermediate states [i.e. band $n$ in Eq.\ (\ref{TPamplitude})],
we can see which transitions do occur, and understand the transfer
of angular momentum.

\subsection{Allowed-forbidden transitions}
When the intermediate state is in the same band as either the
initial or final state (a so-called ``two-band transition''), one
of the photons causes an intraband transition. These two-band
transitions dominate two-photon absorption in GaAs
\cite{vanderZiel77} and indeed in most semiconductors
\cite{Bredikhin73,LeeFan74,Catalano86}. They are
`allowed-forbidden' transitions because the intraband transition,
proportional to the velocity of electrons in the band, is zero at
the $\Gamma$ point. Consequently, it is not possible to derive the
two-photon degree of spin polarization using the states at the
$\Gamma$ point as can be done for one-photon excitation (Fig.\
\ref{Fig_transitions}a) or other two-photon transitions (Fig.\
\ref{Fig_transitions}d). Instead, one must go away from the
$\Gamma $ point and sum over all $\mathbf{k}$ directions. With
this caveat in mind, we nonetheless give a schematic illustration
of a two-band transition in Fig.\ \ref{Fig_transitions}c. One
should bear in mind that, away from the $\Gamma $ point, one
cannot in general associate states in the heavy hole band with
$J_{z}=\pm 3/2$ and states in the light hole band with $J_{z}=\pm
1/2$, since this is only true for $\mathbf{k}
\parallel \mathbf{z}$. It is essentially due to this complication
that the sum over directions of $\mathbf{k}$ gives a two-photon
$P$ that depends on the details of the bands.

The slower decrease of the two-photon $P$ compared to the
one-photon $P$ at excess photon energies greater than the
split-off energy can be understood from a consideration of
two-band transitions. The one-photon $P$ decreases in this regime
due to the selection rules involving transitions from the
split-off band \cite{DP_OpticalOrientation}. The same selection
rules apply to the interband part of the two-band transition, but
the intraband part of transitions from the split-off band is much
weaker than the transitions from the heavy and light hole bands,
since the latter excite to states higher in the conduction band
that have higher velocity.

There are also allowed-forbidden transitions of the three-band
variety ($hh$--$lh$--$c$, $hh$--$so$--$c$, $lh$--$hh$--$c$, and
$lh$--$so$--$c$); in these cases, the inter-valence band matrix
elements can connect states of opposite spin. Their effect on the
two-photon spin polarization approximately cancels out in GaAs, as
one can see by comparing a calculation that neglects them
\cite{BhatSipe00} with one that includes them \cite{Arifzhanov75}.

Within a spherical model, allowed-forbidden transitions must
conserve angular momentum; two-photon absorption with circularly
polarized light must transfer two units of angular momentum to
each electron-hole pair that is created. In order to understand
how this leads to an incomplete spin polarization, one should form
eigenstates of angular momentum, even away from the $\Gamma$
point. Such states can be formed in a spherical model with
envelope functions over an expansion of Bloch states
\cite{Vahala90}. Any treatment of electron angular momentum must
then take into account both the cell-periodic part and the
envelope function part of the electron wavefunction. It is the
latter that is neglected in the argument of Matsuyama
\cite{Matsuyama01}. We plan to return to a more detailed
discussion of this issue in a future publication.

Yet even without that analysis it is clear that, in a simple
two-band spherical model consisting of a single spin degenerate
valence band and a single spin degenerate conduction band, the two
units of angular momentum are divided equally between the two
parts of the electron wavefunction. This can be inferred from the
fact that the envelope function for the relative motion of the
electron and hole has one unit of orbital angular momentum (i.e.
it is a $p$ wave) \cite{Mahan68}. A two-band spherical model can
be mocked-up from an 8 band spherical model by setting the heavy
and light hole band masses equal \cite{Baldereschi70,LeeFan74}.
Doing so with the formula for the two-photon $P$ given by Bhat and
Sipe, \cite{BhatSipe00} one sees that in that case the two-photon
$P$ is 50\% at the band edge. More generally, the maximum
two-photon $P$ in a spherical model is 64\%
\cite{Arifzhanov75,BhatSipe00}.

\subsection{Allowed-allowed transitions}
Allowed-allowed transitions are those for which both matrix
elements in the two-photon amplitude (\ref{TPamplitude}) are
non-zero at the $\Gamma $ point. Allowed-allowed transitions have
a different frequency dependence than allowed-forbidden
transitions. Near $2\hbar \omega \gtrsim E_{g}$ the former varies
as $(2\hbar \omega-E_{g})^{1/2}$ while the latter as $(2\hbar
\omega-E_{g})^{3/2}$. Hence, allowed-allowed transitions can
dominate allowed-forbidden transitions in a frequency range close
to the band edge. For GaAs, however this range is only 10\,meV
\cite{vanderZiel77,HW94}. As seen in the 14 band calculation shown
in the inset of Fig.\ \ref{Fig_DSP_Calc}, the two-photon degree of
spin polarization in this range can be very different from the
rest of the spectrum. These transitions are necessarily due to
lower symmetry parts of the Hamiltonian; in a system with true
spherical symmetry one could not have a two-photon transition from
a $p$ state to an $s$ state, since two-photon transitions cannot
connect states of opposite parity.

The selection rules for allowed-allowed transitions are worked out
in Appendix \ref{sec:allowedallowed}. Consider first the simple
approximation of vanishing interband spin-orbit coupling
$\Delta^{-}$ (denoted $\bar{\Delta}$ in Ref. \onlinecite{PZ96}).
Then the basis states given in Appendix \ref{sec:Notation} are the
energy eigenstates at the $\Gamma$ point. For $\sigma^{+}$
polarized light incident along [001], the only allowed-allowed
transitions are depicted in Fig.\ \ref{Fig_transitions}d; these
can be derived from Table III of Lee and Fan \cite{LeeFan74}. The
product of the two matrix elements in the two-photon amplitude is
the same for both transitions. Thus, if the spin-orbit splitting
of the upper conduction bands $\Delta_{0}^{\prime}$ can be
neglected compared to the other energy differences, then $P$ is
zero [see Eq.\ (\ref{eqn:a_a_dsp_001}) with $\Delta^{-}=0$]. For
$\sigma^{+}$ polarized light incident along [111], the non-zero
transitions are i) $|\Gamma_{8v}^{\prime},+1/2\rangle$ to
$|\Gamma_{8c}^{\prime},-3/2\rangle$ to
$|\Gamma_{6c}^{\prime},-1/2\rangle$; ii)
$|\Gamma_{8v}^{\prime},+3/2\rangle$ to
($|\Gamma_{8c}^{\prime},-1/2\rangle$ and
$|\Gamma_{7c}^{\prime},-1/2\rangle$) to
$|\Gamma_{6c}^{\prime},+1/2\rangle$; and iii)
$|\Gamma_{8v}^{\prime},-3/2\rangle$ to
($|\Gamma_{8c}^{\prime},-1/2\rangle$ and
$|\Gamma_{7c}^{\prime},-1/2\rangle$) to
$|\Gamma_{6c}^{\prime},+1/2\rangle$. Here the prime indicates that
the states are rotated so that the quantization axis is [111]
rather than [001]. If the spin-orbit splitting of the upper
conduction bands $\Delta_{0}^{\prime}$ can be neglected compared
to the other energy differences, then the third of these is zero
and the probability for the second is three times that of the
first, resulting in $P=-0.5$ [see Eq.\ (\ref{eqn:Paa111})].

However, close to the band edge, where allowed-allowed transitions
dominate, the full 14 band calculation (see Table
\ref{Table:bandedgeDSP} or the insets of Figs.\
\ref{Fig_DSP_Calc}--\ref{Fig_ZnSe}) does not agree with these
simple arguments. There is significant difference between
materials; for GaAs $P=-0.21$ and $P=-0.60$ for light incident
along $[001]$ and $[111]$ respectively. The disagreement is due to
the importance of the spin-orbit mixing between the valence and
upper conduction bands, characterized by a nonvanishing
$\Delta^{-}$.

The interband spin-orbit coupling $\Delta^{-}$ would be zero if
the material had inversion symmetry
\cite{Cardona65,Pollak66,CCF88}. In contrast to most of the other
parameters in the 14 band model, the value of $\Delta^{-}$ has not
been determined by directly fitting it to one or more experimental
results. Rather, it has been calculated by various methods: the
empirical pseudopotential method ($-61$\,meV for GaAs)
\cite{Gorczyca91,PZ90,PZ96}, the tight binding method ($-85$\,meV
for GaAs) \cite{CCF88}, the \textit{ab initio}
linear-muffin-tin-orbitals method ($-110$\,meV for GaAs)
\cite{CCF88}, and by an indirect fitting with a $30\times 30$
$\mathbf{k}\cdot \mathbf{p}$ Hamiltonian ($-70$\,meV for GaAs)
\cite{Pollak66, CCF88}.

\begin{figure}
\includegraphics[width=3.1in]{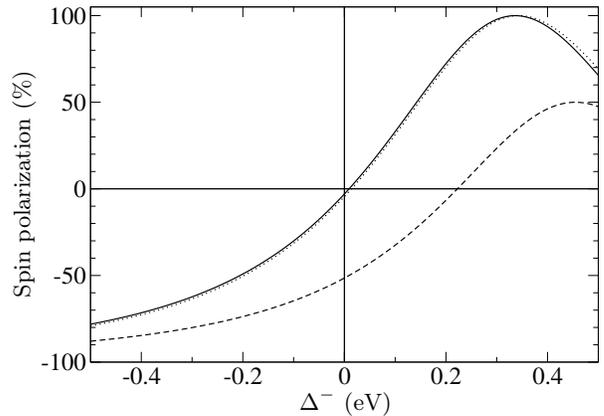}
\caption{Sensitivity of the GaAs band-edge two-photon $P$ to
$\Delta^{-}$. The solid (dashed) line is for two-photon excitation
with circularly polarized light incident along a $\left\langle
001\right\rangle $ ($\left\langle 111\right\rangle $) direction as
calculated with the 14 band model. The dotted line is Eq.\
(\ref{eqn:a_a_dsp_001}).} \label{Fig_VaryDelta}
\end{figure}

In light of the variation in calculated values of the interband
spin-orbit coupling $\Delta^{-}$, we have investigated the
dependence of the band edge two-photon degree of spin polarization
on $\Delta^{-}$. The result, shown in Fig.\ \ref{Fig_VaryDelta},
is rather dramatic. First, it shows that for small $\Delta^{-}$,
$P$ due to [001] incident light is proportional to $\Delta^{-}$,
whereas $P$ due to [111] incident light is less sensitive to
$\Delta^{-}$. Second, it indicates that a 100\% degree of spin
polarization could indeed be possible due to two-photon
absorption. But this possibility is \emph{not} due to the transfer
of angular momentum from the light to the electrons. Since it
results from allowed-allowed transitions that are only non-zero
due to the lack of inversion symmetry and could only occur for
certain crystal orientations, we suggest that some of the angular
momentum comes from the crystal lattice itself.

The selection rules for allowed-allowed transitions including
interband spin-orbit coupling are given for [001] incident light
in Eqs.\
(\ref{eqn:aaSelectionRules1}--\ref{eqn:aaSelectionRules6}) and an
expression for the resulting spin polarization is given in Eq.\
(\ref{eqn:a_a_dsp_001}). It is worth noting that $P$ is
independent of the valence-upper conduction momentum matrix
parameter $Q$.

This allows us to see how the small spin-orbit mixing between
valence and upper conduction bands can have an important effect on
the band edge spin-polarization. Allowed-allowed transitions
between the unmixed states (Fig.\ \ref{Fig_transitions}d) are
proportional to the small matrix element $P_{0}^{\prime}$, which
would be zero if there were inversion symmetry. And since the
intermediate state is in an upper conduction band, the energy
denominator of the two-photon transition amplitude is large, which
further reduces the amplitude for these transitions. The interband
spin-orbit mixing, proportional to $\Delta^{-}/E_{0}^{\prime}$, is
small, but it introduces allowed-allowed transitions with a
valence band as an intermediate state. Then, instead of being
proportional to $P_{0}^{\prime}$, the transition is proportional
to $P_{0}$, and the energy denominator is smaller. This allows the
condition $CP_{0}^{\prime}=DP_{0}\Delta^{-}$ to be met with fairly
modest interband spin-orbit mixing.

\section{Conclusion}

We have shown experimentally that the degrees of spin polarization
produced by one- and two-photon spin injection are approximately
equal in GaAs at an excess photon energy of 90\,meV\@. This was
also recently confirmed experimentally by Stevens \textit{et al}.
\cite{Stevens_pssb}, where for (111)-oriented GaAs the measured
degree of spin polarization of electrons generated by one- and
two-photon excitation was found to be the same, in accord with our
measurements for (001)-oriented GaAs. The experimental results
agree with our theoretical calculations, and they are not at odds
with angular momentum conservation. As well, we have calculated
the degree of spin polarization in other materials to show that
the one- and two-photon degrees of spin polarization need not be
equal.

We have presented the first calculation of two-photon spin
injection that goes beyond a spherical model. The cubic anisotropy
of the two-photon $P$ is small for most of the semiconductors we
investigated at photon energies where allowed-forbidden
transitions dominate, although it is somewhat larger in ZnSe and
InP than in the others. Allowed-allowed transitions, which do not
appear in a spherical model, and hence do not conserve angular
momentum, are found to strongly modify the two-photon $P$ close to
the band edge, and cause a large cubic anisotropy. We have
identified the selection rules responsible for these transitions,
and found that interband spin-orbit coupling plays an important
role.

Measuring the two-photon $P$ due to allowed-allowed transitions
would be challenging in most semiconductors, since they only
dominate in a narrow energy range, and the absorption rate is
small close to the band edge. However, such a measurement could
serve as means of determining the parameter $\Delta^{-}$, which
contributes to the electron $g$ factor \cite{CCF88} and the spin
splitting of bands \cite{MayerRossler93b,PZ96}.

One should bear in mind that Coulomb effects may modify the
two-photon $P$ within an exciton binding energy of the band edge.
The electron spin lifetime will also be shorter within an exciton
binding energy of the band edge due to the Bir-Aronov-Pikus
mechanism of spin relaxation \cite{BAP76}. However, the energy
range where allowed-allowed transitions dominate could in fact be
larger than the exciton binding energy in some materials.

\begin{acknowledgments}
This work was financially supported by the Natural Science and
Engineering Research Council, Photonics Research Ontario, the US
Defense Advanced Research Projects Agency and the US Office of
Naval Research. The authors gratefully acknowledge many
stimulating discussions with Wolfgang R\"{u}hle, Marty Stevens,
Fred Nastos, Ali Najmaie and Eugene Sherman.
\end{acknowledgments}

\appendix
\section{Notation}\label{sec:Notation}
The basis states for the 14 band model are (with $\left| \alpha
_{+}\right\rangle =\left| \uparrow \right\rangle $ and $\left|
\alpha_{-}\right\rangle =\left| \downarrow \right\rangle $),
\begin{eqnarray*}
\left| \Gamma _{7v},\pm 1/2\right\rangle &=& \pm
\frac{1}{\sqrt{3}} \left| Z\right\rangle \left| \alpha _{\pm
}\right\rangle +\frac{1}{\sqrt{3}} \left| X\pm iY\right\rangle
\left| \alpha
_{\mp }\right\rangle \\
 \left| \Gamma _{8v},\pm
1/2\right\rangle &=&\mp \sqrt{\frac{2}{3}} \left| Z\right\rangle
\left| \alpha _{\pm }\right\rangle +\frac{1}{\sqrt{6}} \left| X\pm
iY\right\rangle \left| \alpha _{\mp }\right\rangle \\
 \left| \Gamma _{8v},\pm
3/2\right\rangle &=&\pm \frac{1}{\sqrt{2}} \left| X\pm
iY\right\rangle \left| \alpha _{\pm }\right\rangle
\end{eqnarray*}\begin{eqnarray*} \left| \Gamma _{6c},\pm
1/2\right\rangle &=&i\left| S\right\rangle \left| \alpha _{\pm
}\right\rangle \\ \left| \Gamma _{7c},\pm 1/2\right\rangle &=&\pm
\frac{1}{\sqrt{3}} \left| Z^{\prime }\right\rangle \left| \alpha
_{\pm }\right\rangle +\frac{1}{ \sqrt{3}}\left| X^{\prime }\pm
iY^{\prime }\right\rangle \left| \alpha _{\mp}\right\rangle \\
\left| \Gamma _{8v},\pm 1/2\right\rangle &=&\mp \sqrt{\frac{2}{3}}
\left| Z^{\prime }\right\rangle \left| \alpha _{\pm }\right\rangle
+\frac{1}{ \sqrt{6}}\left| X^{\prime }\pm iY^{\prime}\right\rangle
\left| \alpha _{\mp}\right\rangle \\ \left| \Gamma _{8v},\pm
3/2\right\rangle &=&\pm \frac{1}{\sqrt{2}} \left| X^{\prime }\pm
iY^{\prime }\right\rangle \left| \alpha _{\pm }\right\rangle,
\end{eqnarray*}
where under the point group $T_{d}$, $\left| S\right\rangle $
transforms like $\Gamma_{1}$, while $\left\{ \left| X\right\rangle
,\left| Y\right\rangle ,\left| Z\right\rangle \right\} $ and
$\left\{ \left| X^{\prime }\right\rangle ,\left| Y^{\prime
}\right\rangle ,\left| Z^{\prime }\right\rangle \right\} $
transform like $ \Gamma_{4}$ \cite{PZ90}.

At the $\Gamma $ point, the energy between $\Gamma_{6c}$ and
$\Gamma_{8v}$ bands is $E_{g}$, the energy between $\Gamma_{7c}$
and $\Gamma_{8v}$ bands is $E_{0}^{\prime }$, the energy between
$\Gamma_{8v}$ and $\Gamma_{7v}$ bands is $\Delta _{0}$ and the
energy between $\Gamma_{8c}$ and $\Gamma_{7c}$ bands is
$\Delta_{0}^{\prime }$. The momentum matrix elements are $
P_{0}=i(\hbar /m)\left\langle X\right| p^{x}\left| S\right\rangle
$, $ P_{0}^{\prime }=i(\hbar /m)\left\langle X^{\prime }\right|
p^{x}\left| S\right\rangle $, and $Q=i(\hbar /m)\left\langle
X^{\prime }\right| p^{y}\left| Z\right\rangle $, where $m$ is the
electron mass. The interband spin-orbit coupling is
\begin{equation*}
\Delta ^{-}=\frac{3i\hbar }{4m^{2}c^{2}}\left\langle Z^{\prime
}\right| \left( \mathbf{\nabla }V\times \mathbf{p}\right)
^{y}\left| X\right\rangle ,
\end{equation*}
and its sign has been discussed by Cardona \textit{et al}.
\cite{CCF88}. The parameters $ \gamma_{1L}$, $\gamma_{2L}$, and
$\gamma_{3L}$ are the usual Luttinger parameters that account for
remote band effects on the valence bands. Since the 14 band model
accounts for the $\Gamma_{6c}$, $\Gamma_{7c}$, and $ \Gamma_{8c}$
bands exactly, modified Luttinger parameters are used in the
$14\times 14$ Hamiltonian \cite{PZ96}. The parameter $F$ accounts
for remote band effects on the conduction band ($\Gamma_{6c}$),
essentially fixing its effective mass to the experimentally
observed value. Finally, the parameter $C_{k}$ is the small
$k$-linear term in the valence bands due to interactions with
remote bands \cite{CCF88}.

\section{Allowed-allowed contribution to two-photon spin
injection\label{sec:allowedallowed}}

To calculate $\dot{\mathbf{S}}$ and $\dot{N}$ due to
allowed-allowed transitions, we can approximate all the matrix
elements and energies in the two-photon amplitude by their value
at the $\Gamma $ point, thus avoiding the integral over
$\mathbf{k}$. Since the bands are doubly degenerate at the $\Gamma
$ point, we can use Eq.\ (\ref{Sdot_doublyDengenerate}).

\subsection{Light incident along $\left[ 001\right] $}

Since we use a basis of states with spin quantized along
$\mathbf{\hat{z}}$, $\left\langle c,\Gamma \right|
\hat{S}^{z}\left| c^{\prime },\Gamma \right\rangle \propto
\delta_{c,c^{\prime }}$ and we have
\begin{equation*}
\dot{S}^{z}=\frac{\pi \hbar}{L^{3}} \sum_{v}\left[ \left|
\Omega_{c\uparrow ,v,\Gamma }^{\left( 2\right) }\right|
^{2}-\left| \Omega_{c\downarrow ,v,\Gamma }^{\left( 2\right)
}\right|^{2}\right] \sum_{\mathbf{k}}\delta \left[ 2\omega
-\omega_{cv}\left( \mathbf{k}\right) \right]
\end{equation*}
where $c\uparrow$ and $c\downarrow$ are shorthand for the bands
with states $\left| \Gamma _{6c},\pm 1/2\right\rangle$.

For $\sigma^{+}$ light, with polarization
$\mathbf{\hat{e}}_{\omega}=\left(
\mathbf{\hat{x}}+i\mathbf{\hat{y}}\right) /\sqrt{2}$, and
$\dot{\mathbf{S}}\parallel \mathbf{\hat{z}}$ from Eq.\
(\ref{sDotPhenomCubic}) and the degree of spin polarization is
\begin{equation*}
P=\frac{\sum_{v}\left[
\left|\Omega_{c\downarrow,v,\Gamma}^{\left(2\right)}\right|^{2} -
\left|\Omega_{c\uparrow,v,\Gamma}^{\left(2\right)}\right|^{2}
\right] }{\sum_{v}\left[ \left| \Omega_{c\downarrow ,v,\Gamma
}^{\left( 2\right) }\right|^{2}+\left| \Omega_{c\uparrow ,v,\Gamma
}^{\left( 2\right) }\right|^{2}\right] }.
\end{equation*}

All but the $\Gamma_{6c}$ states are not eigenstates at the
$\Gamma $ point due to spin-orbit coupling between upper
conduction and valence bands parameterized by $\Delta^{-}$. The
Hamiltonian at the $\Gamma$ point in this basis has off-diagonal
elements, but the order of the basis can be arranged so that it is
block diagonal with blocks at most $2\times 2$. For the bands
$\left| \Gamma_{7v},+1/2\right\rangle $ and $\left|
\Gamma_{7c},+1/2\right\rangle $ (or for the bands $\left|
\Gamma_{7v},-1/2\right\rangle $ and $\left|
\Gamma_{7c},-1/2\right\rangle $), the block is
\begin{equation*}
\left[
\begin{array}{ccc}
-E_{g}-\Delta_{0} & & -2\Delta^{-}/3 \\
-2\Delta^{-}/3 &  & E_{0}^{\prime}-E_{g}
\end{array}
\right] .
\end{equation*}
Since $\Delta^{-}/\left( E_{0}^{\prime}+\Delta_{0}\right) \ll 1$,
the off-diagonal part can be treated perturbatively. To first
order in the perturbation, we have eigenvectors
\begin{eqnarray*}
\left| so \uparrow/\downarrow \right\rangle  &=&\left|
\Gamma_{7v},\pm 1/2\right\rangle
+\frac{2\Delta^{-}}{3}\frac{1}{E_{0}^{\prime}+\Delta_{0}}\left|
\Gamma_{7c},\pm 1/2\right\rangle  \\
\left| sc\uparrow/\downarrow \right\rangle  &=&\left|
\Gamma_{7c},\pm 1/2\right\rangle
-\frac{2\Delta^{-}}{3}\frac{1}{E_{0}^{\prime}+\Delta_{0}}\left|
\Gamma_{7v},\pm1/2\right\rangle  .
\end{eqnarray*}
For the $\Gamma_{8}$ bands, the blocks are
\begin{equation*}
\left[
\begin{array}{ccc}
-E_{g} & & \Delta^{-}/3 \\
\Delta^{-}/3 & & E_{0}^{\prime}-E_{g}+\Delta_{0}^{\prime}
\end{array}
\right] ,
\end{equation*}
with eigenvectors to first order in $\Delta^{-}/\left(
E_{0}^{\prime}+\Delta_{0}^{\prime}\right) $,
\begin{eqnarray*}
\left| hh\uparrow/\downarrow \right\rangle  &=& \left|
\Gamma_{8v},\pm 3/2\right\rangle
-\frac{\Delta^{-}}{3}\frac{1}{E_{0}^{\prime}+\Delta_{0}^{\prime}}\left|
\Gamma_{8c},\pm 3/2\right\rangle  \\
\left| lh\uparrow/\downarrow \right\rangle  &=&\left|
\Gamma_{8v},\pm 1/2\right\rangle
-\frac{\Delta^{-}}{3}\frac{1}{E_{0}^{\prime}+\Delta_{0}^{\prime}}\left|
\Gamma_{8c},\pm1/2\right\rangle  \\
\left| hc\uparrow/\downarrow \right\rangle  &=&\left|
\Gamma_{8c},\pm 3/2\right\rangle
+\frac{\Delta^{-}}{3}\frac{1}{E_{0}^{\prime}+\Delta_{0}^{\prime}}\left|
\Gamma_{8v},\pm 3/2\right\rangle  \\
\left| lc\uparrow/\downarrow \right\rangle  &=&\left|
\Gamma_{8c},\pm 1/2\right\rangle
+\frac{\Delta^{-}}{3}\frac{1}{E_{0}^{\prime}+\Delta_{0}^{\prime}}\left|
\Gamma_{8v},\pm 1/2\right\rangle
\end{eqnarray*}

The non-zero matrix elements of $\mathbf{\hat{e}}_{\omega}\cdot
\mathbf{\hat{v}}$ in the eigenstate basis that can cause a
two-photon transition between $v$ and $c$ are
\begin{eqnarray}
\mathbf{e}_{\omega}\cdot \mathbf{v}_{hc\downarrow ,lh\downarrow}
\left( \Gamma \right) &=&-\sqrt{\frac{2}{3}}Q
\label{eqn:aaSelectionRules1}\\ \mathbf{e}_{\omega}\cdot
\mathbf{v}_{c\downarrow,hc\downarrow} \left( \Gamma \right)
&=&P_{0}^{\prime}
+\frac{\Delta^{-}}{3}\frac{1}{E_{0}^{\prime}+\Delta_{0}^{\prime}}P_{0}
\label{eqn:aaSelectionRules2} \\ \mathbf{e}_{\omega}\cdot
\mathbf{v}_{sc\downarrow ,lh\uparrow} \left( \Gamma \right)&=&-Q
\label{eqn:aaSelectionRules3}\\ \mathbf{e}_{\omega}\cdot
\mathbf{v}_{c\uparrow ,sc\downarrow} \left( \Gamma \right)
&=&\sqrt{\frac{2}{3}} \left( P_{0}^{\prime} -
\frac{2\Delta^{-}}{3}\frac{P_{0}}{E_{0}^{\prime}+\Delta_{0}}
\right)
\label{eqn:aaSelectionRules4}\end{eqnarray}\begin{eqnarray}
\mathbf{e}_{\omega}\cdot \mathbf{v}_{so\downarrow ,lh\uparrow}
\left( \Gamma \right)
&=&\frac{-\Delta^{-}Q/3}{E_{0}^{\prime}+\Delta_{0}} \left(
\frac{E_{0}^{\prime}+\Delta_{0}}{E_{0}^{\prime}+\Delta_{0}^{\prime}}
+ 2\right) \label{eqn:aaSelectionRules5}\\
\mathbf{e}_{\omega}\cdot \mathbf{v}_{c\uparrow ,so\downarrow}
\left( \Gamma \right) &=&\sqrt{\frac{2}{3}} \left(
P_{0}+\frac{2\Delta^{-}}{3}\frac{P_{0}^{\prime}}{E_{0}^{\prime} +
\Delta_{0}}\right) ,\label{eqn:aaSelectionRules6}
\end{eqnarray}
where we have dropped terms second order in $\Delta^{-}$. Note
that $\mathbf{e}_{\omega}\cdot \mathbf{v}_{hh\downarrow
,lh\downarrow }\left( \Gamma \right) =0$ by an exact cancellation,
as it should from symmetry considerations. Thus we have
\begin{equation*}
\Omega_{c\downarrow ,lh\downarrow ,\Gamma }^{(2)}=-\left(
\frac{e}{\hbar \omega }\right)^{2}\hbar \left| E_{\omega }\right|
^{2}\sqrt{\frac{2}{3}} Q\left[
AP_{0}^{\prime}+BP_{0}\Delta^{-}\right],
\end{equation*}
where $A \equiv
\left(E_{0}^{\prime}+\Delta_{0}^{\prime}-E_{g}/2\right)^{-1}$ and
$B \equiv
\left(E_{0}^{\prime}+\Delta_{0}^{\prime}\right)^{-1}A/3$. We also
have
\begin{equation*}
\Omega_{c\uparrow ,lh\uparrow ,\Gamma }^{(2)}=-\left(
\frac{e}{\hbar \omega }\right)^{2}\hbar \left| E_{\omega }\right|
^{2}\sqrt{\frac{2}{3}}Q\left[
CP_{0}^{\prime}-DP_{0}\Delta^{-}\right] ,
\end{equation*}
where $C \equiv \left(E_{0}^{\prime}-E_{g}/2\right)^{-1}$,

\begin{equation*}
D \equiv \frac{1}{E_{0}^{\prime}+\Delta_{0}}\left[
\frac{1}{E_{g}/2+\Delta_{0}}\frac{1}{3}\left(
\frac{E_{0}^{\prime}+\Delta_{0}}{E_{0}^{\prime}+\Delta_{0}^{\prime}}
+2\right) +\frac{2}{3}C\right] ,
\end{equation*}
and we have dropped terms proportional to
$QP_{0}^{\prime}(\Delta^{-})^{2}$. The degree of spin polarization
is then
\begin{equation}
P=\frac { \left( AP_{0}^{\prime}+BP_{0} \Delta^{-}\right)^{2} -
\left( CP_{0}^{\prime}-DP_{0}\Delta^{-}\right)^{2} } { \left(
AP_{0}^{\prime}+BP_{0}\Delta^{-}\right)^{2}
+\left(CP_{0}^{\prime}-DP_{0}\Delta^{-}\right)^{2}}.
\label{eqn:a_a_dsp_001}
\end{equation}

\subsection{Light incident along $\left[ 111\right] $}

For $\sigma^{+}$ light incident along $\left[ 111\right] $, it is
more tedious to obtain an expression like Eq.\
(\ref{eqn:a_a_dsp_001}) since there are more non-zero matrix
elements of $\mathbf{\hat{e}}_{\omega}\cdot \mathbf{\hat{v}}$ than
for $\sigma^{+}$ light incident along $\left[ 001\right] $. By
rotating the basis to states quantized along $\left[ 111\right] $,
the matrix of elements of $\mathbf{\hat{e}}_{\omega}\cdot
\mathbf{\hat{v}}$ becomes simpler, but the Hamiltonian is no
longer in $2\times 2$ blocks. When $\Delta^{-}=0$, the latter is
not an issue. In that case, we find
\begin{eqnarray*}
\Omega _{c\downarrow ,lh\uparrow ,\Gamma }^{(2)} &=&\left(
\frac{e}{\hbar \omega }\right) ^{2}\hbar \left| E_{\omega }\right|
^{2}\frac{2}{3}
iQP_{0}^{\prime }A \\
\Omega _{c\uparrow ,hh\uparrow ,\Gamma }^{(2)} &=&\left(
\frac{e}{\hbar \omega }\right) ^{2}\hbar \left| E_{\omega }\right|
^{2}\frac{2}{3}
iQP_{0}^{\prime }\frac{1}{\sqrt{3}}\left( A+2C\right)  \\
\Omega _{c\uparrow ,hh\downarrow ,\Gamma }^{(2)} &=&-\left(
\frac{e}{\hbar \omega }\right) ^{2}\hbar \left| E_{\omega }\right|
^{2}iQP_{0}^{\prime } \frac{1}{3}\sqrt{\frac{2}{3}}
\left(C-A\right) ,
\end{eqnarray*}
where $A$ and $C$ are as defined in the previous subsection, and
$\uparrow $ and $\downarrow $ are along $\left[ 111\right] $. With
the assumption that $\Delta_{0}^{\prime} \ll
E_{0}^{\prime}-E_{g}/2$, $A\approx C$ and we find that
\begin{equation}
    P\left( \Delta^{-}=0\right) =-1/2.\label{eqn:Paa111}
\end{equation}

\end{document}